\documentclass[prl,aps,showpacs,superscriptaddress,tightenlines,twocolumn]{revtex4}
\usepackage{epsfig}
\usepackage[utf8]{inputenc}
\usepackage{float,amsmath,slashed}
\usepackage{graphicx}

\graphicspath{{./}}
\DeclareGraphicsExtensions{.eps}

\newcommand{\beq}{\begin{equation}}
\newcommand{\eeq}{\end{equation}}
\newcommand{\bqa}{\begin{eqnarray}}
\newcommand{\eqa}{\end{eqnarray}}

\def\lsim{\mathrel{\rlap{\lower4pt\hbox{$\sim$}}
    \raise1pt\hbox{$<$}}}                % less than or approx. symbol
\def\gsim{\mathrel{\rlap{\lower4pt\hbox{$\sim$}}
    \raise1pt\hbox{$>$}}}                % greater than or approx. symbol

\begin{document}

\title{Towards a Unified Understanding of Jet Quenching and Elliptic Flow
within Perturbative QCD Parton Transport}
%\title{Towards a common understanding of jet--quenching and elliptic flow within a pQCD description}
%\title{Jet-quenching and elliptic flow --- phenomena of common origin in a partonic transport model}
%\title{A common model for jet--quenching and elliptic flow}
%\title{Jet-quenching and collective flow --- phenomena of common origin in a partonic transport model}
%\title{Description of jet--quenching and elliptic flow in a common model}
%\title{Common origin of jet-quenching and collective flow in a partonic transport model}
%\title{Common origin of jet-quenching and elliptic flow in a partonic transport model}
%\title{Jet--quenching in a partonic transport model}

\author{Oliver Fochler}
\author{Zhe Xu}
\author{Carsten Greiner}
\affiliation{Institut f\"ur Theoretische Physik, Goethe-Universit\"at Frankfurt
am Main \\
  Max-von-Laue-Stra\ss{}e~1,
  D-60438 Frankfurt am Main, Germany\vspace*{2mm}}

\begin{abstract}
The gluonic contribution to the nuclear modification factor $R_{AA}$ is investigated for central Au + Au collisions at $\sqrt{s}
= 200 \mathrm{AGeV}$ employing a perturbative QCD--based parton cascade
including radiative processes. A flat quenching pattern is found up to
transverse momenta of $30\,\mathrm{GeV}$, which is slightly smaller compared with
results from the Gyulassy-Levai-Vitev formalism. We demonstrate that the present microscopic
transport description provides a challenging means of investigating both
jet quenching and a strong buildup of elliptic flow in terms of the same
standard perturbative QCD interactions.
\end{abstract}
\maketitle
\newpage
%%%%%%%%%%%%%%%%%%%%%%%%%%%%%%%%%%%%%%%%%%%%%%%%%%%%%%%%%%%%%%%%%%%%%%%%%%%%%%%%
%\subsection{Introduction}
The phenomenon of jet quenching \cite{Gyulassy:1993hr} has been one
of the two striking discoveries at the Relativistic Heavy Ion
Collider (RHIC) \cite{Adler:2002xw,Adcox:2001jp}. It is widely
believed to provide tomography of the quark-gluon plasma (QGP)
created in ultrarelativistic heavy ion collisions. Furthermore, the
strong elliptic flow, quantified by the Fourier parameter $v_{2}$,
explores the early bulk properties of the medium. Since ideal
hydrodynamics can fairly reproduce the observed $v_{2}$--dependence
on centrality \cite{Huovinen:2001cy}, viscosity in the QGP is
believed to be small \cite{Csernai:2006zz,Lacey:2006bc}. Hence, the
QGP behaves like a perfect fluid, which has been the second major
discovery at RHIC. So far, it has not been possible to relate both
phenomena by a common understanding of the underlying microscopic
processes.

Early attempts to describe the collective behavior by means of standard
binary, elastic perturbative QCD (pQCD) collisions within a parton cascade have
failed
unless (unphysically) large cross sections are employed
\cite{Zhang:1999rs,Lin:2001zk,Molnar:combined}. With such cross sections, as we
will also show in the following, the attenuation of jets
would be far too strong, basically extinguishing all high-momentum
partons. There exist several related theoretical schemes of
the energy loss on the partonic level based on radiative pQCD
interactions that are aimed at describing the observed amount of
jet quenching
\cite{Zakharov:1996fv, Baier:1996sk,Baier:1998yf,Gyulassy:2000er,Jeon:2003gi,
Salgado:2003gb,Wicks:2005gt}.
In particular, the energy loss is found to be dominantly radiative.
It is thus believed that different microscopic physics become
manifest in perturbative interactions of jets with bulk
particles on the one hand, and in seemingly much stronger
interactions between medium constituents that give rise to the liquid-like bulk properties on the other
hand.

It has been recently demonstrated within BAMPS (A Boltzmann approach to multi
parton scatterings) \cite{Xu:2004mz,Xu:2007aa}, a sophisticated
kinetic pQCD based parton cascade {\em including radiative contributions}, that
inelastic Bremsstrahlung
processes can drive the system to equilibration within a short time
less than $1\,\mathrm{fm}/c$. Moreover, pQCD Bremsstrahlung
processes generate a sizeable degree of collective flow
\cite{Xu:2007jv} and a small ratio of shear viscosity to entropy
density $\eta/s$. Using parton-hadron duality the simulated elliptic
flow coefficient $v_2$ has been demonstrated to be in fair agreement
with data employing a coupling constant $\alpha_s = 0.3 \div 0.6$. Some details
on freeze-out \cite{Xu:2008av}, hadronization and possible hadronic phase
contributions have still to be addressed.
For the created gluon plasma $\eta/s$ is in the range $0.08 \div
0.15$, supporting the idea that the medium behaves like a nearly ideal fluid
\cite{Xu:2007ns,Xu:2007jv}.

The present study is motivated by the challenging question whether
the employed treatment of pQCD interactions, including radiative
processes, can simultaneously account for the quenching of
high-momentum partons.

Microscopic transport calculations provide a realistic way of
understanding heavy ion collisions including the full dynamics of the system and allowing for non-thermal initial
conditions. We employ BAMPS to investigate the evolution of gluon
matter produced in heavy ion collisions at RHIC. Treating gluons as
semi-classical and massless Boltzmann particles, elastic $gg
\rightarrow gg$ interactions are included via the leading order pQCD
differential cross section $\frac{d\sigma_{gg\to gg}}{dq_{\perp}^2}
= \frac{9\pi\alpha_{s}^{2}}{(\mathbf{q}_{\perp}^2+m_D^2)^2}$. The
effective Bremsstrahlung matrix element
\begin{eqnarray} \label{eq:M23}
\left|\mathcal{M}_{gg \to ggg}\right|^2 &=& \frac{72 \pi^2
\alpha_s^2 s^2}{(\mathbf{q}_{\perp}^2+m_D^2)^2}\,
 \frac{48 \pi \alpha_s \mathbf{q}_{\perp}^2}
{\mathbf{k}_{\perp}^2
[(\mathbf{k}_{\perp}-\mathbf{q}_{\perp})^2+m_D^2]}\times\nonumber\\
&&\times\Theta\left( \frac{\Lambda_g}{\gamma} - \tau \right)
\end{eqnarray}
is used for inelastic gluon multiplication and annihilation
processes. $\mathbf{q}_{\perp}$ and $\mathbf{k}_{\perp}$ denote the
perpendicular components of the momentum transfer and of the
radiated gluon momentum in the center of mass (CM) frame of the colliding
particles, respectively. Gluon interactions are screened by a dynamically computed Debye
mass $m_D^2 = d_G \pi \alpha_s \int \frac{d^3p}{(2\pi)^3}
\frac{1}{p} N_c f$, where $d_G=16$ is the gluon degeneracy factor
for $N_c=3$ and $f=f(p,x,t)$ is the local distribution function.

In the treatment of Bremsstrahlung processes the coherent Landau,
Pomeranchuk and Migdal (LPM) effect \cite{Migdal:1956tc}, leads to a
suppression of the emission rate for high energy particles. Such interference effects cannot be included in the
microscopic transport calculations. Therefore, the theta function in
(\ref{eq:M23}) is introduced to ensure that only independent
processes are considered, i.e. that only processes within the so
called Bethe-Heitler regime are taken into consideration. Physically
this implies that the formation time $\tau$ of the emitted gluon
must not exceed the mean free path of the parent particle. The mean
free path $\Lambda_{g}$ is given in the local comoving frame of a
computational spatial cell, which is not identical with the frame
where the gluon is emitted purely transversal and thus $\tau =
1/k_{\perp}$. A Lorentz factor $\gamma = \frac{\cosh
y}{\sqrt{1-\beta^2}}\,(1 + \beta\, \tanh y\, \cos\theta)$ has to be
employed. $\beta$ denotes the boost velocity from the comoving frame
to the CM frame of the colliding particles, $y$ is the
rapidity of the emitted gluon measured from the CM frame and
$\theta$ is the angle between $\vec{\beta}$ and the axis of the
colliding particles in the CM frame. For thermal energies
the boost $\beta$ becomes small and the theta function reduces to
$\Theta(k_{\perp}\Lambda_g - \cosh y)$.

\begin{figure}[htb]
  \begin{center}
    \includegraphics[width=7.4cm]{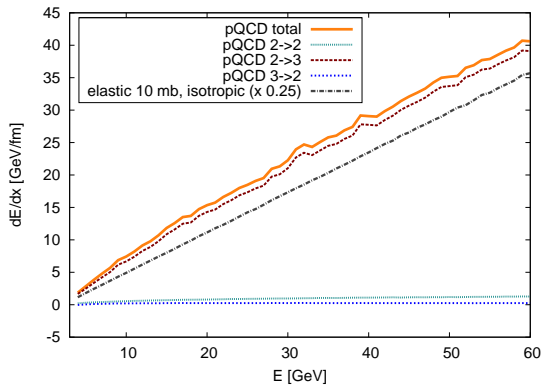}
    \caption{(Color online) Differential energy loss of a gluon jet in a static
and thermal
    medium of gluons with $T=400\,\mathrm{MeV}$.
    The contributions of the different pQCD processes to the total $dE/dx$
    are shown. Additionally $dE/dx$ scaled down by a factor of $4$ is shown for
a gluon jet
    interacting only via isotropic binary scatterings with a fixed cross
    section $\sigma = 10\,\mathrm{mb}$.}
    \label{fig:dEdx}
  \end{center}
\vspace{-0.5cm}
\end{figure}

The cutoff in (\ref{eq:M23}) affects the overall value of the
interaction probability for $gg \leftrightarrow ggg$ as well as the momentum
shape of the spectrum of the radiated gluons. Fig. \ref{fig:dEdx} shows the
differential energy loss $dE/dx$ of a high energy gluon jet in a
static and thermal medium of gluons with temperature
$T=400\,\mathrm{MeV}$ as a function of the jet energy $E$. Throughout this work
we employ a fixed coupling constant $\alpha_s=0.3$. The elastic energy loss from $gg
\rightarrow gg$ interactions is found
to be only slowly rising with energy to a level of $dE/dx \approx
1\,\mathrm{GeV}/{fm}$ at $E=40\,\mathrm{GeV}$ and as expected the contribution
from $ggg \rightarrow gg$ processes is negligible with $dE/dx \approx
0.3\,\mathrm{GeV}/{\mathrm{fm}}$. Radiative $gg \rightarrow ggg$ processes
strongly dominate the
energy loss, for instance resulting in a total $dE/dx \approx
29\,\mathrm{GeV}/{\mathrm{fm}}$ at $E = 40\,\mathrm{GeV}$, with $dE/dx$
almost linearly increasing with the jet energy $E$. In comparison to the Arnold-Moore-Yaffe formalism \cite{Jeon:2003gi} a hard gluon with $E = 40\,\mathrm{GeV}$ loses $75\,\%$ of its energy about a factor of 2 faster \cite{Schenke:2009priv}.
 
The individual cross sections moderatly increase with the jet energy, yielding a total $\left< \sigma_{\mathrm{tot}} \right> \approx 3.5\,\mathrm{mb}$ at $E = 40\,\mathrm{GeV}$ \cite{FXG:2009bb} for a medium as characterized above. We note that a dependence of $dE/dx$ on the distance the parton has propagated cannot be observed in our approach, since coherent effects are explicitely not included.

For arguments to come, Fig. \ref{fig:dEdx} also shows $dE/dx$
(scaled down by a factor of $4$) for a gluon jet interacting purely
elastic with the medium at a fixed cross section of
$\sigma = 10\,\mathrm{mb}$ and an isotropic angular distribution.
The differential energy loss at this temperature is about a factor of $3 \div 4$
larger than $dE/dx$ from pQCD
interactions, e.g. $\left.dE/dx\right|_{\sigma=10\,\mathrm{mb}}
\approx 94\,\mathrm{GeV}/\mathrm{fm}$ at $E=40\,\mathrm{GeV}$.
%%%%%%%%%%%%%%%%%%%%%%%%%%%%%%%%%%%%%%%%%%%%%%%%%%%%%%%%%%%%%%%%%%%%%%%%%%%%%%

%\subsection{Thermalization and collective flow}
The initial gluon distribution for Au+Au collisions at 200~AGeV is
chosen to be an ensemble of mini-jets with a lower momentum cut-off
$p_{0} = 1.4\,\mathrm{GeV}$, produced in nucleon-nucleon collisions
following a Glauber-model with a Wood-Saxon density profile and
Gl\"uck-Reya-Vogt parton distribution functions \cite{Gluck:1994uf}
with a $K$--factor of $2$. Free streaming is applied to regions
where the energy density has dropped below a critical energy density
of $\varepsilon_{c} = 1\, \mathrm{GeV}/\mathrm{fm}^3$. With this setup it has been demonstrated
\cite{Xu:2007aa,Xu:2007jv,Xu:2008av} that the experimental findings for the
rapidity
distribution of transverse energy and the flow parameter for various
centralities are nicely reproduced.

\begin{figure}[htb]
  \begin{center}
    \includegraphics[width=7.4cm]{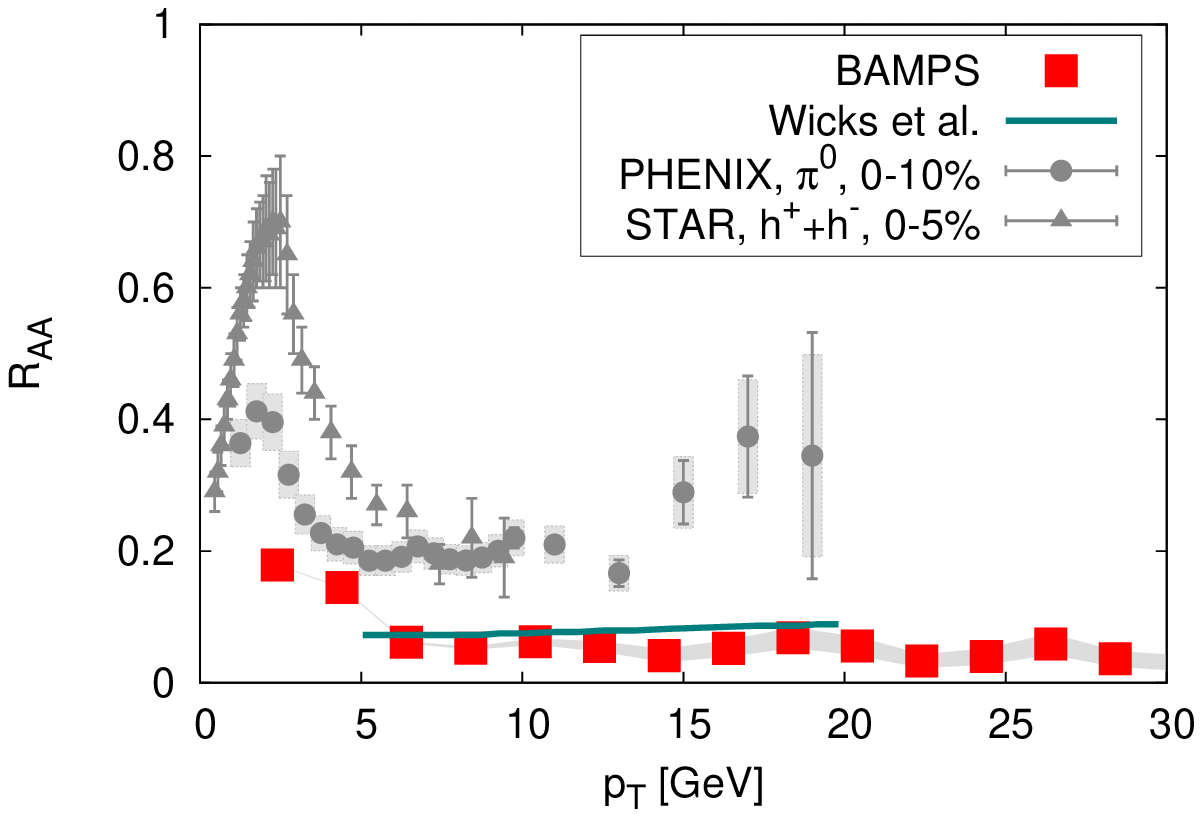}
    \vspace{-0.5cm}
    \includegraphics[width=7.4cm]{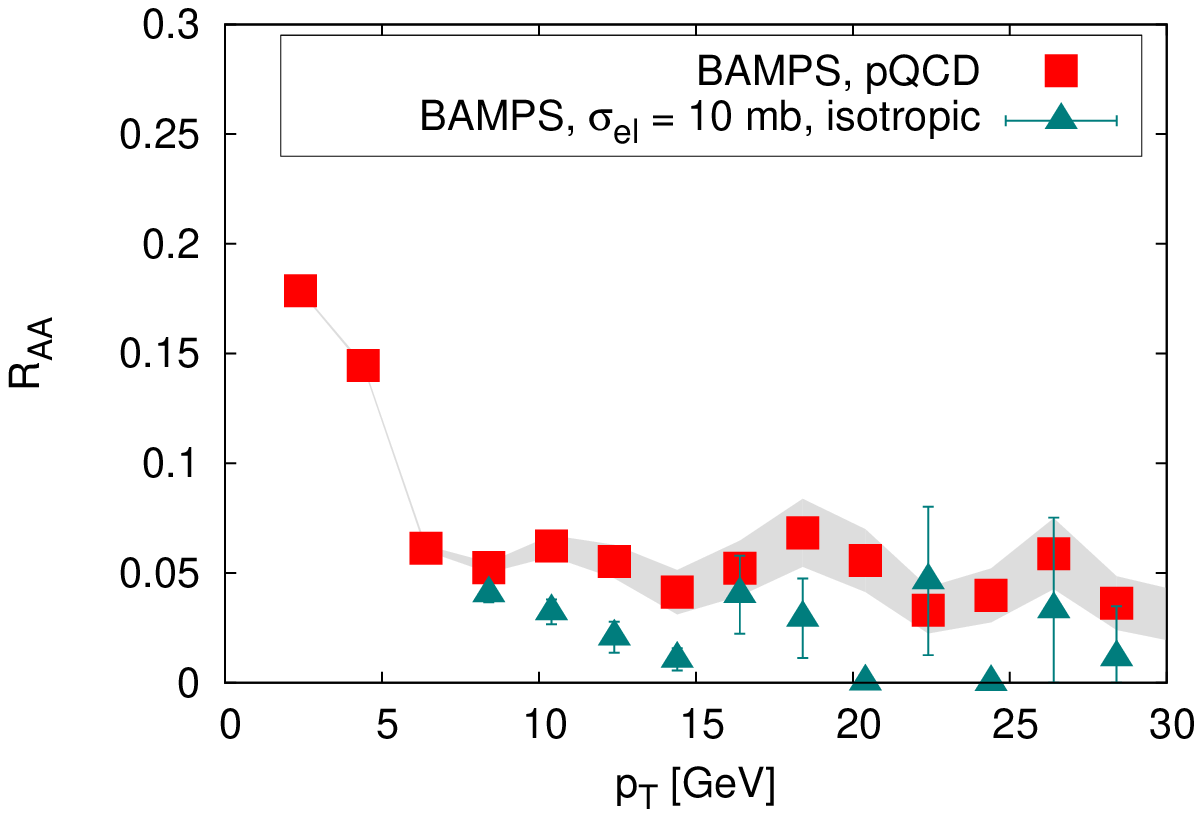}
    \caption{(Color online) Upper: Gluonic $R_{AA}$ at midrapidity ($y \,\epsilon\, [-0.5,0.5]$) as extracted from simulations for central Au+Au collisions at 200~AGeV. The shaded area indicates the statistical error. For direct comparison the result from Wicks et al. \cite{Wicks:2005gt} for the gluonic contribution to $R_{AA}$ and experimental results from PHENIX \cite{Adare:2008qa} for $\pi^{0}$ and STAR \cite{Adams:2003kv} for charged hadrons are shown. Lower: Gluonic $R_{AA}$ (triangles) for a scenario where all particles with $p_{T} > 8\,\mathrm{GeV}$ can only interact elastically with fixed $\sigma = 10\,\mathrm{mb}$.}
    \label{fig:RAA}
  \end{center}
\vspace{-0.5cm}
\end{figure}
Due to the steeply falling momentum spectrum of the initial jets one
would need to simulate an infeasibly high number of events in order
to obtain sufficient statistics at high--$p_{T}$. We have therefore developed a
suitable weighting and reconstruction scheme. The
underlying idea is to simulate a huge number of {\em initial
spectra}, which is rather straightforward from the computational
point of view, and to select such events for further simulations
that contain high-$p_{T}$ partons. The results then need to be
appropriately weighted. For this we characterize each initial state
according to $X = \max(p_{T,i})$, the maximum $p_{T}$ in a given
rapidity range. From each bin $j$ in $X$ a number of $N_j$ events is
simulated, the results are averaged within these bins and finally
combined with appropriate weights~$P_j$. This procedure has been
thoroughly tested \cite{FXG:2009bb} and, so far, allows for the
investigation of observables up to $p_{T}\approx 30\,\mathrm{GeV}$.
In the following we use a bin-size of 1~GeV and select up to
$N_j=120$ events per bin for full simulation.

Jet quenching is generally specified in terms of the nuclear
modification factor
\begin{equation} \label{eq:RAA}
R_{AA}=\frac{d^{2}N_{AA}/dp_{T}dy}{T_{AA}d^{2}\sigma_{NN}/dp_{T}dy}\text{,}
\end{equation}
which is the ratio of the particle yield in a A+A collision at given
$p_{T}$ and $y$ to the yield in a p+p collision scaled by the
appropriate number of binary collisions. We directly compute
$R_{AA}$ by taking the ratio of the final $p_{T}$ spectra to the
initial mini-jet spectra. In this Letter we concentrate on central
($b=0\,\mathrm{fm}$) collisions; $R_{AA}$ for non-central collisions
can be straightforwardly studied and will be presented in an
upcoming work \cite{FXG:2009bb}. Fig. \ref{fig:RAA} shows the result
for the gluonic contribution to $R_{AA}$, exhibiting a clear
suppression of high--$p_{T}$ gluon jets at a roughly constant level
of $R_{AA}^{\mathrm{gluons}} \approx 0.053$. This constitutes the major result
of the present study.

The suppression of gluon jets is approximately a factor $3 \div 4$ stronger than
seen in experimental pion data. An excessive quenching, however, was to be
expected since at present the simulation does not include quarks, which are
bound to lose considerably less energy due to their color factor and dominate the initially produced jets from $p_{T} \approx 20\,\mathrm{GeV}$. As also discussed in \cite{Xu:2008av}, the inclusion of light quark degrees of freedom would
yield better agreement between BAMPS results and data on particle spectra,
mean-$p_{T}$ and the $p_{T}$-dependence of elliptic flow.
Indeed, comparing with state of the art results from Wicks et al.
\cite{Wicks:2005gt}, obtained within the GLV formalism \cite{Gyulassy:2000er}, for the gluonic
contribution to $R_{AA}$ (seen as the line in Fig. \ref{fig:RAA}), which in
their approach together with the quark contribution reproduces the experimental
data, one finds better agreement. The gluonic suppression in their calculations
is found to be $R_{AA} \approx 0.07 \div 0.09$. They assume a gluon density per
unit rapidity of $dN_g/dy=1100$ and the strong expansion of the fireball has to
be included via an effective description. In the present microscopic treatment
the
full dynamics are selfconsistently included, where the initial gluon density is
given by $dN_g/dy \approx 700$, which then evolves to a final $dN_g/dy\approx
800$. The inclusion of light quarks will provide important means of further
verification and will be addressed in a future work.

\begin{figure}[htb]
  \begin{center}
    \includegraphics[width=7.4cm]{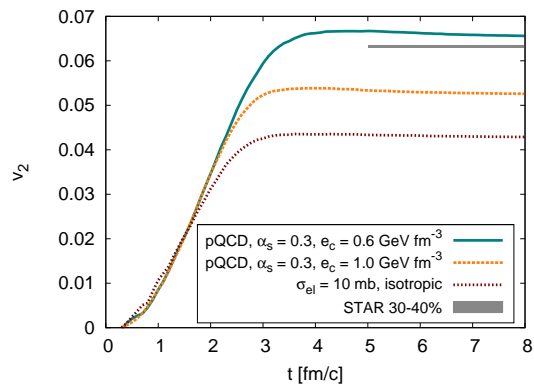}
    \caption{(Color online) Generation of elliptic flow in BAMPS at midrapidity
for a noncentral collision with an impact parameter of
$b=8.6\,\mathrm{fm}$, obtained from calculations with
$\alpha_{s}=0.3$ and two freeze-out energy densities,
$\varepsilon_{c}=0.6$~and~$1.0\,\mathrm{GeV}\,\mathrm{fm}^{-3}$
\cite{Xu:2008av}.
    For comparison the result employing solely binary collisions with constant
    $\sigma = 10\,\mathrm{mb}$
($\varepsilon_{c}=1.0\,\mathrm{GeV}\,\mathrm{fm}^{-3}$) is shown. The
grey-shaded band represents the
    experimental value \cite{Adams:2004bi}.}
    \label{fig:v2t}
  \end{center}
\vspace{-0.5cm}
\end{figure}

To demonstrate the importance of inelastic processes for explaining
both elliptic flow and jet quenching within a combined framework, in Fig.
\ref{fig:v2t} we compare the elliptic flow for a scenario of only
binary and isotropic collisions with $\sigma = 10\,\mathrm{mb}$ to
the full results from \cite{Xu:2008av}. Though the cross section is
large, it still does not fully succeed in explaining the strong
elliptic flow. A still larger one has to be (phenomenologically) employed in
order to generate enough flow \cite{Molnar:combined}. On the other
hand, we also investigate a scenario in which jet-like gluons with
$p_{T}>8\,\mathrm{GeV}$ can only interact elastically with the same fixed and
isotropic cross section of $\sigma = 10\,\mathrm{mb}$. In this calculation, the
medium particles interact with the usual pQCD based matrix elements,
including inelastic collisions. As already expected from Fig. \ref{fig:dEdx},
the results shown in the lower panel of Fig. \ref{fig:RAA} explicitly
demonstrate that already with $\sigma = 10\,\mathrm{mb}$ the suppression is too
strong. So, while approaches using fixed and large cross sections might be able
to reproduce thermalization times, angular correlations
\cite{Ma:2006rn} or elliptic
flow \cite{Molnar:combined,Zhang:2005ni}, they would fail at
simultaneously reproducing the correct nuclear modification factor.

\begin{figure}[htb]
  \begin{center}
    \includegraphics[width=6.5cm]{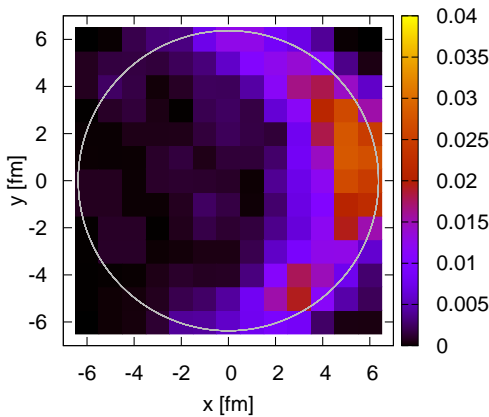}
    \caption{(Color online) Probability distribution of production points of
jets with final $p_{T}>10\,\mathrm{GeV}$.
     Events are rotated such that jets in the final state travel in the positive
$x$--direction.}
    \label{fig:jetOrigin}
  \end{center}
\vspace{-0.5cm}
\end{figure}

%%%%%%%%%%%%%%%%%%%%%%%%%%%%%%%%%%%%%%%%%%%%%%%%%%%%%%%%%%%%%%%%%%%%%%%%%%%%%%%
%\subsection{Jet origin}
The use of a full microscopic transport treatment also offers the
possibility to investigate observables, that are not directly
accessible in experiment.
Fig. \ref{fig:jetOrigin} shows the probability distribution of production points
in the transverse
plane of gluon jets with a final $p_{T} > 10\,\mathrm{GeV}$ in a
central ($b=0\,\mathrm{fm}$) Au+Au collision using the full pQCD
BAMPS version. Events in this illustration are rotated such that the
jets emerge with their final $\vec{p}_{T}$ parallel to the
$x$--axis and directed in positive $x$--direction. The spatial
extent of the initial production points is roughly controlled by the
parameter $R_{A}$ in the Wood--Saxon distribution $n_{A}(r)=n_0 /
(1+e^{(r-R_A)/d})$, with $R_A=6.37\,\mathrm{fm}$ in our case. A bias
towards production points near the surface is visible as would be
naively expected for the strong jet suppression seen in Fig.
\ref{fig:RAA}. The escaping jets have been produced on the average $1.7\,\mathrm{fm}$ away from the surface and have undergone $\left< N_{23}
\right> \approx 0.6$ radiative processes ($\left< N_{22} \right> \approx 2.0$
and $\left< N_{32} \right> \approx 0.1$). Still, at such high transverse momenta
a few jets can indeed probe the interior of the reaction volume, corroborating
the notion that the inclusion of the full medium dynamics is crucial to the
investigation of jet quenching.

%%%%%%%%%%%%%%%%%%%%%%%%%%%%%%%%%%%%%%%%%%%%%%%%%%%%%%%%%%%%%%%%%%%%%%%%%%%%%%%
%\subsection{Conclusions}
For the first time a consistent and fully pQCD based microscopic parton
transport description has been applied as a common setup to both elliptic flow
and jet quenching observed in RHIC experiments. As established previously, the
gluon matter simulated in the parton cascade BAMPS exhibits a sizeable pressure
build--up and a small ratio of shear viscosity to entropy, $\eta/s$. In this
work we found that the suppression of high--$p_{T}$ gluon jets is roughly
constant at $R_{AA}^{\mathrm{gluons}} \approx 0.053$ over a large $p_{T}$ range,
with inelastic $gg \leftrightarrow ggg$ processes being the main cause of
energy loss. The observed level of jet quenching is in reasonable agreement
with recent analytic results, though the suppression of gluonic jets appears to
be somewhat stronger. Future studies including light quarks and a
fragmentation scheme into the simulation will help to further confirm our
findings that bulk properties of the medium and the energy loss of
high--$p_{T}$ partons can be described within a common transport
approach. The investigation of elliptic flow at large transverse momenta will
also provide important means of cross--checking.

%%%%%%%%%%%%%%%%%%%%%%%%%%%%%%%%%%%%%%%%%%%%%%%%%%%%%%%%%%%%%%%%%%%%%%%%%%%%%%%

%%%%%%%%%%%%%%%%%%%%%%%%%%%%%%%%%%%%%%%%%%%%%%%%%%%%%%%%%%%%%%%%%%%%%%%%%%%%%%%
The authors would like to thank M. Gyulassy for stimulating
discussions throughout this work. The simulations were performed at the Center
for Scientific Computing of the Goethe University Frankfurt.

\bibliography{fochler}

\begin{thebibliography}{30}
\expandafter\ifx\csname natexlab\endcsname\relax\def\natexlab#1{#1}\fi
\expandafter\ifx\csname bibnamefont\endcsname\relax
  \def\bibnamefont#1{#1}\fi
\expandafter\ifx\csname bibfnamefont\endcsname\relax
  \def\bibfnamefont#1{#1}\fi
\expandafter\ifx\csname citenamefont\endcsname\relax
  \def\citenamefont#1{#1}\fi
\expandafter\ifx\csname url\endcsname\relax
  \def\url#1{\texttt{#1}}\fi
\expandafter\ifx\csname urlprefix\endcsname\relax\def\urlprefix{URL }\fi
\providecommand{\bibinfo}[2]{#2}
\providecommand{\eprint}[2][]{\url{#2}}

\bibitem[{\citenamefont{Gyulassy and Wang}(1994)}]{Gyulassy:1993hr}
\bibinfo{author}{\bibfnamefont{M.}~\bibnamefont{Gyulassy}} \bibnamefont{and}
  \bibinfo{author}{\bibfnamefont{X.-N.} \bibnamefont{Wang}},
  \bibinfo{journal}{Nucl. Phys.} \textbf{\bibinfo{volume}{B420}},
  \bibinfo{pages}{583} (\bibinfo{year}{1994}).

\bibitem[{\citenamefont{Adler et~al.}(2002)}]{Adler:2002xw}
\bibinfo{author}{\bibfnamefont{C.}~\bibnamefont{Adler}} \bibnamefont{et~al.}
  (\bibinfo{collaboration}{STAR}), \bibinfo{journal}{Phys. Rev. Lett.}
  \textbf{\bibinfo{volume}{89}}, \bibinfo{pages}{202301}
  (\bibinfo{year}{2002}).

\bibitem[{\citenamefont{Adcox et~al.}(2002)}]{Adcox:2001jp}
\bibinfo{author}{\bibfnamefont{K.}~\bibnamefont{Adcox}} \bibnamefont{et~al.}
  (\bibinfo{collaboration}{PHENIX}), \bibinfo{journal}{Phys. Rev. Lett.}
  \textbf{\bibinfo{volume}{88}}, \bibinfo{pages}{022301}
  (\bibinfo{year}{2002}).

\bibitem[{\citenamefont{Huovinen et~al.}(2001)\citenamefont{Huovinen, Kolb,
  Heinz, Ruuskanen, and Voloshin}}]{Huovinen:2001cy}
\bibinfo{author}{\bibfnamefont{P.}~\bibnamefont{Huovinen}},
  \bibinfo{author}{\bibfnamefont{P.~F.} \bibnamefont{Kolb}},
  \bibinfo{author}{\bibfnamefont{U.~W.} \bibnamefont{Heinz}},
  \bibinfo{author}{\bibfnamefont{P.~V.} \bibnamefont{Ruuskanen}},
  \bibnamefont{and} \bibinfo{author}{\bibfnamefont{S.~A.}
  \bibnamefont{Voloshin}}, \bibinfo{journal}{Phys. Lett.}
  \textbf{\bibinfo{volume}{B503}}, \bibinfo{pages}{58} (\bibinfo{year}{2001}),
  \eprint{hep-ph/0101136}.

\bibitem[{\citenamefont{Csernai et~al.}(2006)\citenamefont{Csernai, Kapusta,
  and McLerran}}]{Csernai:2006zz}
\bibinfo{author}{\bibfnamefont{L.~P.} \bibnamefont{Csernai}},
  \bibinfo{author}{\bibfnamefont{J.~I.} \bibnamefont{Kapusta}},
  \bibnamefont{and} \bibinfo{author}{\bibfnamefont{L.~D.}
  \bibnamefont{McLerran}}, \bibinfo{journal}{Phys. Rev. Lett.}
  \textbf{\bibinfo{volume}{97}}, \bibinfo{pages}{152303}
  (\bibinfo{year}{2006}).

\bibitem[{\citenamefont{Lacey et~al.}(2007)}]{Lacey:2006bc}
\bibinfo{author}{\bibfnamefont{R.~A.} \bibnamefont{Lacey}}
  \bibnamefont{et~al.}, \bibinfo{journal}{Phys. Rev. Lett.}
  \textbf{\bibinfo{volume}{98}}, \bibinfo{pages}{092301}
  (\bibinfo{year}{2007}).

\bibitem[{\citenamefont{Zhang et~al.}(1999)\citenamefont{Zhang, Gyulassy, and
  Ko}}]{Zhang:1999rs}
\bibinfo{author}{\bibfnamefont{B.}~\bibnamefont{Zhang}},
  \bibinfo{author}{\bibfnamefont{M.}~\bibnamefont{Gyulassy}}, \bibnamefont{and}
  \bibinfo{author}{\bibfnamefont{C.~M.} \bibnamefont{Ko}},
  \bibinfo{journal}{Phys. Lett.} \textbf{\bibinfo{volume}{B455}},
  \bibinfo{pages}{45} (\bibinfo{year}{1999}).

\bibitem[{\citenamefont{Lin and Ko}(2002)}]{Lin:2001zk}
\bibinfo{author}{\bibfnamefont{Z.-w.} \bibnamefont{Lin}} \bibnamefont{and}
  \bibinfo{author}{\bibfnamefont{C.~M.} \bibnamefont{Ko}},
  \bibinfo{journal}{Phys. Rev.} \textbf{\bibinfo{volume}{C65}},
  \bibinfo{pages}{034904} (\bibinfo{year}{2002}).

\bibitem[{\citenamefont{Molnar and Gyulassy}(2002)}]{Molnar:combined}
\bibinfo{author}{\bibfnamefont{D.}~\bibnamefont{Molnar}} \bibnamefont{and}
  \bibinfo{author}{\bibfnamefont{M.}~\bibnamefont{Gyulassy}},
  \bibinfo{journal}{Nucl. Phys.} \textbf{\bibinfo{volume}{A697}},
  \bibinfo{pages}{495} (\bibinfo{year}{2002}); \bibinfo{author}{\bibfnamefont{D.}~\bibnamefont{Molnar}} \bibnamefont{and}
  \bibinfo{author}{\bibfnamefont{P.}~\bibnamefont{Huovinen}},
  \bibinfo{journal}{Phys. Rev. Lett.} \textbf{\bibinfo{volume}{94}},
  \bibinfo{pages}{012302} (\bibinfo{year}{2005}).

\bibitem[{\citenamefont{Zakharov}(1996)}]{Zakharov:1996fv}
\bibinfo{author}{\bibfnamefont{B.~G.} \bibnamefont{Zakharov}},
  \bibinfo{journal}{JETP Lett.} \textbf{\bibinfo{volume}{63}},
  \bibinfo{pages}{952} (\bibinfo{year}{1996}).

\bibitem[{\citenamefont{Baier et~al.}(1997)\citenamefont{Baier, Dokshitzer,
  Mueller, Peigne, and Schiff}}]{Baier:1996sk}
\bibinfo{author}{\bibfnamefont{R.}~\bibnamefont{Baier}},
  \bibinfo{author}{\bibfnamefont{Y.~L.} \bibnamefont{Dokshitzer}},
  \bibinfo{author}{\bibfnamefont{A.~H.} \bibnamefont{Mueller}},
  \bibinfo{author}{\bibfnamefont{S.}~\bibnamefont{Peigne}}, \bibnamefont{and}
  \bibinfo{author}{\bibfnamefont{D.}~\bibnamefont{Schiff}},
  \bibinfo{journal}{Nucl. Phys.} \textbf{\bibinfo{volume}{B484}},
  \bibinfo{pages}{265} (\bibinfo{year}{1997}).

\bibitem[{\citenamefont{Baier et~al.}(1998)\citenamefont{Baier, Dokshitzer,
  Mueller, and Schiff}}]{Baier:1998yf}
\bibinfo{author}{\bibfnamefont{R.}~\bibnamefont{Baier}},
  \bibinfo{author}{\bibfnamefont{Y.~L.} \bibnamefont{Dokshitzer}},
  \bibinfo{author}{\bibfnamefont{A.~H.} \bibnamefont{Mueller}},
  \bibnamefont{and} \bibinfo{author}{\bibfnamefont{D.}~\bibnamefont{Schiff}},
  \bibinfo{journal}{Phys. Rev.} \textbf{\bibinfo{volume}{C58}},
  \bibinfo{pages}{1706} (\bibinfo{year}{1998}).

\bibitem[{\citenamefont{Gyulassy et~al.}(2001)\citenamefont{Gyulassy, Levai,
  and Vitev}}]{Gyulassy:2000er}
\bibinfo{author}{\bibfnamefont{M.}~\bibnamefont{Gyulassy}},
  \bibinfo{author}{\bibfnamefont{P.}~\bibnamefont{Levai}}, \bibnamefont{and}
  \bibinfo{author}{\bibfnamefont{I.}~\bibnamefont{Vitev}},
  \bibinfo{journal}{Nucl. Phys.} \textbf{\bibinfo{volume}{B594}},
  \bibinfo{pages}{371} (\bibinfo{year}{2001}).

\bibitem[{\citenamefont{Jeon and Moore}(2005)}]{Jeon:2003gi}
\bibinfo{author}{\bibfnamefont{S.}~\bibnamefont{Jeon}} \bibnamefont{and}
  \bibinfo{author}{\bibfnamefont{G.~D.} \bibnamefont{Moore}},
  \bibinfo{journal}{Phys. Rev.} \textbf{\bibinfo{volume}{C71}},
  \bibinfo{pages}{034901} (\bibinfo{year}{2005}).

\bibitem[{\citenamefont{Salgado and Wiedemann}(2003)}]{Salgado:2003gb}
\bibinfo{author}{\bibfnamefont{C.~A.} \bibnamefont{Salgado}} \bibnamefont{and}
  \bibinfo{author}{\bibfnamefont{U.~A.} \bibnamefont{Wiedemann}},
  \bibinfo{journal}{Phys. Rev.} \textbf{\bibinfo{volume}{D68}},
  \bibinfo{pages}{014008} (\bibinfo{year}{2003}).

\bibitem[{\citenamefont{Wicks et~al.}(2007)\citenamefont{Wicks, Horowitz,
  Djordjevic, and Gyulassy}}]{Wicks:2005gt}
\bibinfo{author}{\bibfnamefont{S.}~\bibnamefont{Wicks}},
  \bibinfo{author}{\bibfnamefont{W.}~\bibnamefont{Horowitz}},
  \bibinfo{author}{\bibfnamefont{M.}~\bibnamefont{Djordjevic}},
  \bibnamefont{and} \bibinfo{author}{\bibfnamefont{M.}~\bibnamefont{Gyulassy}},
  \bibinfo{journal}{Nucl. Phys.} \textbf{\bibinfo{volume}{A784}},
  \bibinfo{pages}{426} (\bibinfo{year}{2007}).

\bibitem[{\citenamefont{Xu and Greiner}(2005)}]{Xu:2004mz}
\bibinfo{author}{\bibfnamefont{Z.}~\bibnamefont{Xu}} \bibnamefont{and}
  \bibinfo{author}{\bibfnamefont{C.}~\bibnamefont{Greiner}},
  \bibinfo{journal}{Phys. Rev.} \textbf{\bibinfo{volume}{C71}},
  \bibinfo{pages}{064901} (\bibinfo{year}{2005}).

\bibitem[{\citenamefont{Xu and Greiner}(2007)}]{Xu:2007aa}
\bibinfo{author}{\bibfnamefont{Z.}~\bibnamefont{Xu}} \bibnamefont{and}
  \bibinfo{author}{\bibfnamefont{C.}~\bibnamefont{Greiner}},
  \bibinfo{journal}{Phys. Rev.} \textbf{\bibinfo{volume}{C76}},
  \bibinfo{pages}{024911} (\bibinfo{year}{2007}).

\bibitem[{\citenamefont{Xu et~al.}(2008)\citenamefont{Xu, Greiner, and
  Stocker}}]{Xu:2007jv}
\bibinfo{author}{\bibfnamefont{Z.}~\bibnamefont{Xu}},
  \bibinfo{author}{\bibfnamefont{C.}~\bibnamefont{Greiner}}, \bibnamefont{and}
  \bibinfo{author}{\bibfnamefont{H.}~\bibnamefont{Stocker}},
  \bibinfo{journal}{Phys. Rev. Lett.} \textbf{\bibinfo{volume}{101}},
  \bibinfo{pages}{082302} (\bibinfo{year}{2008}).

\bibitem[{\citenamefont{Xu and Greiner}(2009)}]{Xu:2008av}
\bibinfo{author}{\bibfnamefont{Z.}~\bibnamefont{Xu}} \bibnamefont{and}
  \bibinfo{author}{\bibfnamefont{C.}~\bibnamefont{Greiner}},
  \bibinfo{journal}{Phys. Rev.} \textbf{\bibinfo{volume}{C79}},
  \bibinfo{pages}{014904} (\bibinfo{year}{2009}).

\bibitem[{\citenamefont{Xu and Greiner}(2008{\natexlab{b}})}]{Xu:2007ns}
\bibinfo{author}{\bibfnamefont{Z.}~\bibnamefont{Xu}} \bibnamefont{and}
  \bibinfo{author}{\bibfnamefont{C.}~\bibnamefont{Greiner}},
  \bibinfo{journal}{Phys. Rev. Lett.} \textbf{\bibinfo{volume}{100}},
  \bibinfo{pages}{172301} (\bibinfo{year}{2008}{\natexlab{b}}),
  \eprint{0710.5719}.

\bibitem[{\citenamefont{Migdal}(1956)}]{Migdal:1956tc}
\bibinfo{author}{\bibfnamefont{A.~B.} \bibnamefont{Migdal}},
  \bibinfo{journal}{Phys. Rev.} \textbf{\bibinfo{volume}{103}},
  \bibinfo{pages}{1811} (\bibinfo{year}{1956}).

\bibitem[{\citenamefont{Schenke}(2009)}]{Schenke:2009priv}
\bibinfo{author}{\bibfnamefont{B.}~\bibnamefont{Schenke}},
  \bibinfo{journal}{priv. comm.}  (\bibinfo{year}{2009}).

\bibitem[{\citenamefont{Fochler et~al.}(2009)\citenamefont{Fochler, Xu, and
  Greiner}}]{FXG:2009bb}
\bibinfo{author}{\bibfnamefont{O.}~\bibnamefont{Fochler}},
  \bibinfo{author}{\bibfnamefont{Z.}~\bibnamefont{Xu}}, \bibnamefont{and}
  \bibinfo{author}{\bibfnamefont{C.}~\bibnamefont{Greiner}},
  \bibinfo{journal}{forthcoming}  (\bibinfo{year}{2009}).

\bibitem[{\citenamefont{Gluck et~al.}(1995)\citenamefont{Gluck, Reya, and
  Vogt}}]{Gluck:1994uf}
\bibinfo{author}{\bibfnamefont{M.}~\bibnamefont{Gluck}},
  \bibinfo{author}{\bibfnamefont{E.}~\bibnamefont{Reya}}, \bibnamefont{and}
  \bibinfo{author}{\bibfnamefont{A.}~\bibnamefont{Vogt}}, \bibinfo{journal}{Z.
  Phys.} \textbf{\bibinfo{volume}{C67}}, \bibinfo{pages}{433}
  (\bibinfo{year}{1995}).

\bibitem[{\citenamefont{Ma et~al.}(2007)}]{Ma:2006rn}
\bibinfo{author}{\bibfnamefont{G.~L.} \bibnamefont{Ma}} \bibnamefont{et~al.},
  \bibinfo{journal}{Phys. Lett.} \textbf{\bibinfo{volume}{B647}},
  \bibinfo{pages}{122} (\bibinfo{year}{2007}).

\bibitem[{\citenamefont{Zhang et~al.}(2005)\citenamefont{Zhang, Chen, and
  Ko}}]{Zhang:2005ni}
\bibinfo{author}{\bibfnamefont{B.}~\bibnamefont{Zhang}},
  \bibinfo{author}{\bibfnamefont{L.-W.} \bibnamefont{Chen}}, \bibnamefont{and}
  \bibinfo{author}{\bibfnamefont{C.-M.} \bibnamefont{Ko}},
  \bibinfo{journal}{Phys. Rev.} \textbf{\bibinfo{volume}{C72}},
  \bibinfo{pages}{024906} (\bibinfo{year}{2005}).

\bibitem[{\citenamefont{Adare et~al.}(2008)}]{Adare:2008qa}
\bibinfo{author}{\bibfnamefont{A.}~\bibnamefont{Adare}} \bibnamefont{et~al.}
  (\bibinfo{collaboration}{PHENIX}), \bibinfo{journal}{Phys. Rev. Lett.}
  \textbf{\bibinfo{volume}{101}}, \bibinfo{pages}{232301}
  (\bibinfo{year}{2008}).

\bibitem[{\citenamefont{Adams et~al.}(2003)}]{Adams:2003kv}
\bibinfo{author}{\bibfnamefont{J.}~\bibnamefont{Adams}} \bibnamefont{et~al.}
  (\bibinfo{collaboration}{STAR}), \bibinfo{journal}{Phys. Rev. Lett.}
  \textbf{\bibinfo{volume}{91}}, \bibinfo{pages}{172302}
  (\bibinfo{year}{2003}).

\bibitem[{\citenamefont{Adams et~al.}(2005)}]{Adams:2004bi}
\bibinfo{author}{\bibfnamefont{J.}~\bibnamefont{Adams}} \bibnamefont{et~al.}
  (\bibinfo{collaboration}{STAR}), \bibinfo{journal}{Phys. Rev.}
  \textbf{\bibinfo{volume}{C72}}, \bibinfo{pages}{014904}
  (\bibinfo{year}{2005}).

\end{thebibliography}

\end{document}